# Electric double layer in concentrated solutions of ionic surfactants

Roumen Tsekov

Department of Physical Chemistry, University of Sofia, 1164 Sofia, Bulgaria

A simple non-local theoretical model is developed considering concentrated ionic surfactant solutions as regular ones. Their thermodynamics is described by the Cahn-Hilliard theory coupled with electrostatics. It is discovered that unstable solutions possess two critical temperatures, where the temperature coefficients of all characteristic lengths are discontinuous. At temperatures below the lower critical temperature ionic surfactant solutions separate into thin layers of oppositely charged liquids spread across the whole system and the electric potential is strictly periodic. At temperatures between the two critical temperatures separation can occur only near the solution surface thus leading to an oscillatory-decaying electric double layer. At temperatures above the higher critical temperature as well as in stable solutions there is no separation and the electric potential decays exponentially.

The theory of the electric double layer [1] dates back to the classical works of Helmholtz, Gouy, Chapman, Stern, Debye, Hückel and others. Due to general mathematical complications in dense systems, however, the applications are mainly restricted to dilute ionic solutions, where the electric potential is described via the Poisson-Boltzmann equation. The interactions between electric double layers are also intensively studied as an important component of the inter-particle and colloidal forces [2]. Recently, significant attention has been attracted by highly charged Coulomb mixtures, where many specific phenomena take place [3] among them mono-species electric double layers [4]. General theories of charged fluids are also developed [5-8], which account for ion correlations going beyond the Poisson-Boltzmann theory. An interesting aspect here is the effect of non-electrostatic interactions between the ions in the electric double layer expected to become important in concentrated solutions. The aim of the present paper is to develop a simple theoretical approach to concentrated ionic surfactant solutions based on electrostatics coupled with the regular solution model. The latter is described in accordance to the Cahn-Hilliard theory [9]. It is shown that a temperature region exists where the electric potential is an oscillatory-decaying function. At temperatures below this temperature region the spatial arrangement of the ions is purely periodic thus indicating a complete phase separation to oppositely charged layers. On the contrary, at temperatures above this temperature region the oscillations are completely damped and the electric potential decays exponentially. On the borders of this temperature region the temperature coefficients of all characteristic lengths exhibit singularities, thus indicating a new type of second-order phase-transitions.

The structure of concentrated solutions is rather complex and, for this reason, it is difficult for theoretical modeling. Hereafter a simple model is developed, which aims to describe the main features of the electric double layer in concentrated solutions of symmetric ionic surfactants. The latter dissociate in water completely to cations and anions with charges $\pm z$, respectively. The electric potential $\phi$ is related to the local charge density via the Poisson equation

$$\varepsilon_0 \varepsilon \Delta \phi = ze(c_- - c_+) \tag{1}$$

where $\varepsilon_0 \varepsilon$ is the dielectric permittivity of the solution, $\Delta$ is the Laplace operator and $c_-$ and $c_+$ are the local concentrations of anions and cations. The further application requires a relation between the local ion concentrations and the electric potential. In the frames of the regular solution model the electrochemical potentials of the ions can be presented in the forms

$$\tilde{\mu}_+ = \mu_+^0 + k_B T \ln c_+ + ze\phi + \omega c_-^2 / 2c_\infty^2 - \lambda \Delta c_+ / 2c_\infty \tag{2}$$

$$\tilde{\mu}_- = \mu_-^0 + k_B T \ln c_- - ze\phi + \omega c_+^2 / 2c_\infty^2 - \lambda \Delta c_- / 2c_\infty \tag{3}$$

Here $T$ is temperature, $c_\infty$ is the in the bulk concentration far away from the surface, $\omega$ is the interaction energy excess between ions and the last terms follow from the Cahn-Hilliard gradient theory [9]. Note that the non-negative parameter $\lambda$ is proportional to the surface tension between the two possible oppositely charged solutions dominantly containing cations and anions, respectively. Since these two charged liquids contribute equally to the surface tension on their common dividing surface, the elastic constant $\lambda$ is the same in both electrochemical potentials above. The presence of the last terms in Eqs. (2) and (3) is an indication of non-locality because the electrochemical potential depends not only on the local concentration but also on the concentration distribution all over the solution. The Cahn-Hilliard theory is an extension to mixtures of the general van der Waals density functional theory of non-homogeneous fluids [10]. It is a first approximation for non-locality but a very general one accounting for the leading effects of the ion correlations. There are no restrictions of its application due to high concentrations. On the contrary, the Cahn-Hilliard theory is usually applied to condensed phases since in dilute systems $\lambda$ is negligibly small. If the explicit potentials of the ion interactions are known it is a matter of integration [9] only to calculate rigorously $\omega$ and $\lambda$.

The thermodynamic equilibrium requires constant values of the electrochemical potentials along the solution, which can be determined from Eqs. (2) and (3) far away from the surface, where $c_- = c_+ = c_\infty$ and $\phi = 0$, i.e.

$$\tilde{\mu}_+ = \mu_+^0 + k_B T \ln c_\infty + \omega/2 \qquad\qquad \tilde{\mu}_- = \mu_-^0 + k_B T \ln c_\infty + \omega/2 \qquad (4)$$

Introducing these expressions back into Eqs. (2) and (3), respectively, leads to the following expression for the electric potential

$$2ze\phi = k_B T \ln(c_-/c_+) - \omega(c_-^2 - c_+^2)/2c_\infty^2 - \lambda \Delta(c_- - c_+)/2c_\infty \qquad (5)$$

Due to the strong electrostatic effects the local concentrations differ slightly from $c_\infty$ and the first entropic and second enthalpic terms in Eq. (5) can be linearized to obtain

$$2zec_\infty \phi = (k_B T - \omega)(c_- - c_+) - \lambda \Delta(c_- - c_+)/2 \qquad (6)$$

Finally, substituting the local concentration variations in Eq. (6) by Eq. (1) leads to the following differential equations for the electric potential $\phi$

$$-\beta\lambda\Delta\Delta\phi + 2(1-\beta\omega)\Delta\phi = 2\kappa^2\phi \qquad (7)$$

Here $\beta \equiv 1/k_B T$ and $\kappa \equiv (2z^2 e^2 c_\infty / \varepsilon_0 \varepsilon k_B T)^{1/2}$ are the reciprocal thermal energy and Debye length, respectively. Note that Eq. (7) reduces to the linearized Poisson-Boltzmann equation from the Debye-Hückel theory in the case of dilute solutions, where $\omega = 0$ and $\lambda = 0$. Due to the first non-local term, however, it goes beyond the local screening theories being proven to be inaccurate in concentrated ionic solutions.

Deep inside the solution the molar fraction of cations and anions are equal due to electro-neutrality. Close to the surface, however, the specific adsorption of surfactant ions induces decrease of surfactant ions concentration and increase of the counterions one. To analyze the spatial patterns in the electric double layer close to the surface it is more convenient to apply the Fourier transform of Eq. (7) which reflects into the following dispersion relation

$$\beta\lambda q^4 + 2(1-\beta\omega)q^2 + 2\kappa^2 = 0 \qquad (8)$$

for the wave vector $q$. In general Eq. (8) has four roots

$$q = \pm\sqrt{\frac{\pm\sqrt{(1-\beta\omega)^2 - 2\beta\lambda\kappa^2} - (1-\beta\omega)}{\beta\lambda}} \qquad (9)$$

Physically relevant roots are, however, ones with non-negative imaginary parts since the Fourier components with $q_{Im} < 0$ diverge at infinity. Since the real parts of the physically relevant roots appear with opposite signs $\pm q_{Re}$, the dependence of the electric potential near a flat surface acquires the form

$$\phi = \phi_s \exp(q_{Im} z) \cos(q_{Re} z) \tag{10}$$

where $\phi_s$ is the surface potential and the solution occupies the lower half-space $z \leq 0$. The boundary conditions of Eq. (7) to obtain Eq. (10) are a fixed surface potential and purely exponential decay very near to the surface due to the requirement of space needed for building of another layer similarly charged to the surface.

If $\omega \leq 0$, the solution is stable at any temperature. Since in this case the cation-anion attraction is stronger than the cation-cation and anion-anion ones, the surface tension between the possible oppositely charged liquids should vanish. Hence, in this case one can accept generally $\lambda = 0$ and Eq. (9) reduces to

$$q = i\kappa / \sqrt{1 - \beta\omega} \tag{11}$$

As seen, the wave vector is purely imaginary thus indicating simple exponential decay of the potential. The decay length $[\varepsilon_0 \varepsilon (k_B T - \omega)/2z^2 e^2 c_\infty]^{1/2}$, however, is larger than the Debye one due to the ion interactions, which obviously help the entropy to diffuse the electric double layer.

More interesting is the case $\omega > 0$, where the solution can separate in two oppositely charged liquids at low temperatures. This case is especially relevant to surfactant solutions since the surfactant ions are usually complex organic species. Hence, the strong non-electrostatic attraction between them as well as the hydrophobic repulsion between surfactant and strongly hydrated counter ions can result in a positive $\omega$. If the material parameters $\omega$, $\lambda$, $c_\infty$ and $\varepsilon$ are temperature independent constants, one can introduce the following characteristic temperatures: the classical critical temperature $T_c \equiv \omega / k_B$ and the split temperature $T_s \equiv 2(\lambda z^2 e^2 c_\infty / \varepsilon_0 \varepsilon)^{1/2} / k_B$. Alternatively the latter can be presented as $T_s = 2\lambda \kappa_s^2 / k_B$, where $\kappa_s \equiv \kappa(T_s) = (z^2 e^2 c_\infty / \varepsilon_0 \varepsilon \lambda)^{1/4}$. Hence, $k_B T_s$ is proportional to the mean energy of the electrostatic contribution to the surface tension between the oppositely charged liquids. Using these notations Eq. (9) can be rewritten as

$$q/\kappa_s = \pm\sqrt{2(\pm\sqrt{t^2 - 1} - t)} \tag{12}$$

where $t \equiv (T - T_c)/T_s$. The Cahn-Hilliard parameter $\lambda \sim \omega a^2 = k_B T_c a^2$ is proportional to interaction energy $\omega$ and a length $a$, reflecting the typical distance of the intermolecular interactions. Hence, introducing $\kappa_c \equiv \kappa(T_c)$ the ratio between the split and critical temperatures scales as $T_s/T_c \sim \kappa_c a$, while $\kappa_s \sim (\kappa_c/a)^{1/2}$. At $c_\infty = 0.1$ M, $z = 1$ and a room critical temperature these estimates lead to $\kappa_s \sim \kappa_c = 1$ nm$^{-1}$ and $T_s/T_c \sim 0.1$ at $a \sim 0.1$ nm.

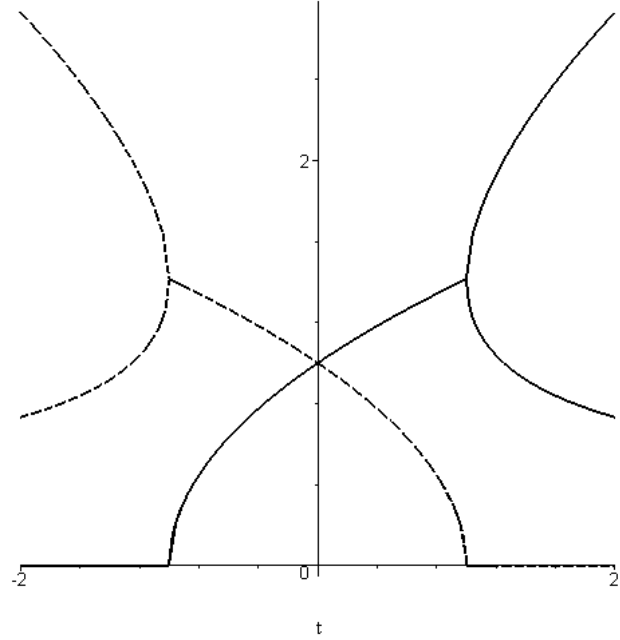

**Fig. 1** The universal temperature dependence of the dimensionless components of the wave vector $q_{Re}/\kappa_s$ (dashed line) and $q_{Im}/\kappa_s$ (solid line) on $t \equiv (T - T_c)/T_s$

In Fig. 1 the corresponding universal dependence of the physically relevant real and imaginary parts of the dimensionless wave vector $q/\kappa_s$ are presented as functions of the dimensionless temperature difference $t$. As seen for temperatures below $T_{cb} \equiv T_c - T_s$ the system decomposes of layers of oppositely charged liquids, which spread out periodically across the whole liquid. In this case $q_{Im} = 0$ and $q_{Re}$ is inverse proportional to the thickness of the layers. Since the upper branch corresponds to extremely thin layers, a realistic structure will be described by the lower branch $q_{Re}^2 \leq 2\kappa_s^2$. Such layers are a particular example of the lamellar structures formed by surfactants at high concentrations. For temperature above $T_{cb}$ but bellow $T_{cs} \equiv T_c + T_s$ both imaginary $q_{Im} = \kappa_s(1+t)^{1/2}$ and real $q_{Re} = \kappa_s(1-t)^{1/2}$ parts are not zero thus indicating an oscillatory-decaying propagation of the electric potential as described in Eq. (10). In this case the liquid separates to oppositely charged layers only near the surface. At $T \geq T_{cs}$, however, $q_{\text{Re}}=0$ and the potential becomes purely exponentially decaying but its decay length differs from the Debye one. Here again the realistic value is $q_{\text{Im}}^2 \leq 2\kappa_s^2$ since at large temperatures it tends asymptotically to $\kappa$. In general the symmetry rule $q_{\text{Im}}(t)=q_{\text{Re}}(-t)$ holds. At both new critical temperatures, the bulk one $T_{cb}$ and the surface one $T_{cs}$, the temperature coefficients of $q_{\text{Re}}$ and $q_{\text{Im}}$ are discontinuous, while at $T=T_c$ the real and imaginary parts are both equal to $\kappa_s$. Note that a decrease of the ionic surfactant concentration reflects in decrease of $T_s$. Hence, in relatively dilute solutions both the new critical temperatures tend to the classical one $T_c$.

The present non-local model of concentrated ionic surfactant solutions as regular ones is a simple and effective tool for description and understanding of the complex structure of the electric double layer. Essential innovation in addition of the Cahn-Hilliard gradient term is the coupling with electrostatics. The main limitation of the theory is a linearization but even so it demonstrates interesting and important results. In general, the described phase transitions are related to spinodal decomposition of the ionic surfactant solutions to oppositely charged liquids. However, in contrast to simple non-charged liquids it is favorable for the ions to form thin layers instead of bulk phases since the coexisting phases are charged solutions. The interplay between thermodynamic instability and electrostatics leads also to appearance of two critical temperatures spanning a temperature region, where the decomposition is only possible near the solution surface. Such boundary induced phase-transition generates an oscillatory-decaying electric double layer. Similar oscillatory-decaying dependence of the electrical potential is observed in room temperature ionic liquids via computer simulations [11-13]. Moreover, singularities of the interfacial polarization are experimentally detected in ionic liquids [14-16], which are explained also by demixing criticality [17]. An interesting question remained open is what happens in highly concentrated solutions without non-electrostatic interactions between ions. Intuitively one would expect that $\omega<0$ due to electrostatic repulsion and attraction between similar and dissimilar charges, respectively. Computer simulations [18] have shown, however, that due to strong correlation effects purely Coulomb mixtures possess also critical temperatures, under which phase separation takes place.